\documentclass[12pt]{article}

\usepackage{amsmath}
\allowdisplaybreaks

\usepackage{amssymb}

\begin{document}

\baselineskip=16.8pt plus 0.2pt minus 0.1pt

\makeatletter
\@addtoreset{equation}{section}
\renewcommand{\theequation}{\thesection.\arabic{equation}}
\renewcommand{\thefootnote}{\fnsymbol{footnote}}

\newcommand{\p}{\partial}
\newcommand{\nn}{\nonumber}
\newcommand{\ds}{\displaystyle}
\newcommand{\Pmatrix}[1]{\begin{pmatrix} #1 \end{pmatrix}}
\newcommand{\bm}[1]{\boldsymbol{#1}}
\newcommand{\bra}[1]{\left\langle #1\right\vert}
\newcommand{\ket}[1]{\left\vert #1\right\rangle}
\newcommand{\braket}[2]{\langle #1\vert #2\rangle}
\newcommand{\wh}[1]{\widehat{#1}}
\newcommand{\Drv}[2]{\frac{\p #1}{\p #2}}
\newcommand{\abs}[1]{\left\vert #1\right\vert}
\newcommand{\T}{{\rm T}}
\newcommand{\calN}{{\cal N}}
\renewcommand{\Re}{\mathop{{\rm Re}}}
\renewcommand{\Im}{\mathop{{\rm Im}}}
\newcommand{\Res}{\mathop{{\rm Res}}}
\newcommand{\Omgb}{\ket{\Omega_b}}
\newcommand{\SQb}{\ket{\Xi_b}}
\newcommand{\Sb}{S'}
\newcommand{\Tb}{T'}
\newcommand{\Cp}{C'}
\newcommand{\Vzz}{V_{00}}
\newcommand{\drv}[2]{\frac{d #1}{d #2}}
\newcommand{\Mz}{M_0}
\newcommand{\Mzb}{\Mz'}
\newcommand{\eps}{\epsilon}
\newcommand{\ooint}[2]{\oint\limits_{#2}\!\frac{d #1}{2\pi i}}
\newcommand{\arctanh}{\mathop{\rm arctanh}}
\newcommand{\tz}{t}
\newcommand{\Muk}{\ket{D_{\tz}}}
\newcommand{\Mukp}[1]{\bigl|\widetilde{D}_{\tz}(#1)\bigr\rangle}
\newcommand{\Mukpof}{\ket{D_{\pi/4}}}
\newcommand{\mL}{\ell}
\newcommand{\mB}{\lambda}
\newcommand{\fugo}{\epsilon}
\newcommand{\z}{\xi}
\newcommand{\y}{\pi}
\newcommand{\bmzb}{\bm{\z}'}
\newcommand{\bmyb}{\bm{\y}'}
\newcommand{\newz}{\Xi}
\newcommand{\newy}{\Pi}
\newcommand{\newbmz}{\bm{\newz}}
\newcommand{\newbmy}{\bm{\newy}}
\newcommand{\ketproj}{\ket{\mbox{Proj}}}

\begin{titlepage}
\title{
\hfill\parbox{4cm}
{\normalsize KUNS-1966\\CALT-68-2555\\{\tt hep-th/0504184}}\\
\vspace{1cm}
{\bf Boundary and Midpoint Behaviors of Lump Solutions
in Vacuum String Field Theory
}
}
\author{
Hiroyuki {\sc Hata}
\thanks{{\tt hata@gauge.scphys.kyoto-u.ac.jp}}\\
{\sl Department of Physics, Kyoto University,
Kyoto 606-8502, Japan}\\[5pt]
and\\[5pt]
Sanefumi {\sc Moriyama}
\thanks{{\tt moriyama@math.nagoya-u.ac.jp}}
\\
{\sl California Institute of Technology 452-48,
Pasadena, CA91125, USA}\\
and\\
{\sl Graduate School of Mathematics, Nagoya University}\\
{\sl Nagoya 464-8602, Japan}
\thanks{Permanent address}
\\[15pt]
}
\date{\normalsize April, 2005}
\maketitle
\thispagestyle{empty}
\begin{abstract}
\normalsize

We discuss various issues concerning the behaviors near the
boundary ($\sigma=0,\pi$) and the midpoint ($\sigma=\pi/2$) of the
open string coordinate $X(\sigma)$ and its conjugate momentum
$P(\sigma)=-i\delta/\delta X(\sigma)$ acting on the matter
projectors of vacuum string field theory.
Our original interest is in the dynamical change of the boundary
conditions of the open string coordinate from the Neumann one in the
translationally invariant backgrounds to the Dirichlet one in the
D-brane backgrounds.
We find that the Dirichlet boundary condition is realized on a lump
solution only partially and only when its parameter takes a special
value. On the other hand, the string midpoint has a mysterious
property: it obeys the Neumann (Dirichlet) condition in the
translationally invariant (lump) background.

\end{abstract}

\end{titlepage}

\section{Introduction}
\label{sec:Intro}

One of the most celebrated catch-phrases for string field theory (SFT)
is the background independence.
Although perturbative string theory is constructed on a particular
background, SFT is expected to connect different backgrounds by
reexpanding the string field around the classical solution
representing the new background.
Hence, we expect that, with SFT we are able to understand how physics
around the background changes dynamically.

An ideal laboratory for this interesting mechanism is the open string
sector. According to Sen \cite{descent}, the tachyon in a
certain open string theory with D-branes indicates the instability
of the D-brane. After the tachyon condensates, we arrive at states
of lower dimensional D-branes and finally at the true vacuum
without D-branes (tachyon vacuum).
In the SFT description \cite{tachyon}, the theories before and after
the tachyon condensation, namely, the theories with D-branes of
various dimensions and the one without any D-branes are connected
simply by the shift of the string field $\Phi$; $\Phi\to\Phi_C +\Phi$
with $\Phi_C$ being a classical solution of SFT. In this sense, the
action of SFT is common among the backgrounds.

This background independence of open SFT, however, leads to an
apparent contradiction as we shall explain (let us consider the
bosonic open string theory in the following).
The open string field $\Phi[X^\mu(\sigma),c(\sigma),b(\sigma)]$ is a
functional of the space-time string coordinate $X^\mu(\sigma)$ as well
as the (anti-)ghost coordinates, $c(\sigma)$ and $b(\sigma)$.
Before the tachyon condensation (i.e., in open SFT in the presence of
a space-filling D25-brane), the string coordinate $X^\mu(\sigma)$ and
its conjugate momentum $P_\mu(\sigma)=-i\delta/\delta X^\mu(\sigma)$
are subject to the Neumann boundary condition (BC) for all
$\mu=0,1,\cdots,25$:
\begin{equation}
\drv{X^\mu(\sigma)}{\sigma}\biggr|_{\sigma=0,\pi}=
\drv{P_\mu(\sigma)}{\sigma}\biggr|_{\sigma=0,\pi}=0 .
\label{eq:NBC}
\end{equation}
Therefore, they are expanded in term of the cosine modes:
\begin{align}
X^\mu(\sigma)&
=\wh{x}^\mu+i\sum_{n=1}^\infty
\left(a_n - a_n^\dagger\right)^\mu \z_n(\sigma) ,
\label{eq:X}
\\
P_\mu(\sigma)&
=\frac{1}{\pi}\!\left[\wh{p}_\mu +
\sum_{n=1}^\infty\left(a_n + a_n^\dagger\right)_\mu
\y_n(\sigma)\right] ,
\label{eq:P}
\end{align}
where $\z_n(\sigma)$ and $\y_n(\sigma)$ are defined by
\begin{align}
\z_n(\sigma)&=\sqrt{\frac{2}{n}}\,\cos n\sigma ,
\label{eq:z_n}
\\
\y_n(\sigma)&=\sqrt{\frac{n}{2}}\,\cos n\sigma ,
\label{eq:y_n}
\end{align}
and $(\wh{x},\wh{p})$ and $(a_n,a_n^\dagger)$ satisfy
\begin{equation}
\left[\wh{x}^\mu,\wh{p}_\nu\right]=i\delta^\mu{}_\nu,
\qquad
\left[a_m^\mu,a_n^{\nu\dagger}\right]=\delta_{m,n}\,\eta^{\mu\nu}.
\end{equation}
On the other hand, in open SFT in the background of lower dimensional
D-branes, the string coordinates $X^\perp(\sigma)$ perpendicular to the
D-brane should obey the Dirichlet BC:
\begin{equation}
X^\perp(\sigma=0,\pi)=a^\perp ,
\label{eq:DBC}
\end{equation}
where $a^\perp$ denotes the perpendicular coordinates of the brane.
However, the two boundary conditions \eqref{eq:NBC} and \eqref{eq:DBC}
are inconsistent if the two open string fields $\Phi^{(25)}$ and
$\Phi^{(p)}$ describing the fluctuations around a D25-brane and a
D$p$-brane with $p<25$ are related through the shift
$\Phi^{(25)}=\Phi_C+\Phi^{(p)}$ using the D$p$-brane classical
solution $\Phi_C$ in the SFT around the D25-brane.

One way to resolve this puzzle is to start with the open SFT of Witten
\cite{CSFT} describing the D25-brane background and study the SFT
obtained by reexpanding around the classical lump solution
representing D$p$-branes with $p<25$.
Although the classical lump solutions in the open SFT have been
obtained in the level truncation approximation to give
expected results for their energy density \cite{HarKra,MJMT,MSZ},
it seems hard to give a definite answer to our question of the change
of the boundary conditions within this approximation.
In this paper we shall adopt another way: we start with vacuum
string field theory (VSFT) \cite{VSFT,solutions}, which is a candidate
SFT expanded around the tachyon vacuum without any D-branes. The
space-time open string coordinate $X^\mu(\sigma)$ as an argument of
the string field of VSFT and its conjugate $P_\mu(\sigma)$ are subject
to the Neumann BC and have expansions \eqref{eq:X} and \eqref{eq:P}
since VSFT should describe the translationally invariant tachyon
vacuum. What is good about VSFT is that exact lump solutions have
been constructed \cite{solutions,BCFT,Muk}.
Therefore, it is expected that we can carry out
exact analyses on how the boundary condition switches to the Dirichlet
one for the string coordinates perpendicular to these lumps.

Then, in what sense can the boundary condition change from the Neumann
one to the Dirichlet one in the lump solution background of VSFT?
The most conservative and modest test is whether the physical
excitation spectrum around a lump solution agrees with that on the
D-brane. Here we would like to pursue another possibility:
$(X^\perp(\sigma),P_\perp(\sigma))$ acting on any
fluctuation modes around a lump solution satisfy the Dirichlet BC
due to some singular nature of the solution
\cite{singular,anomaly,exact}.
Recall that the matter part $\Psi_C$ of the $p$-dimensional lump
solution in VSFT satisfying the projector condition
\begin{equation}
\Psi_C *\Psi_C=\Psi_C ,
\label{eq:Psi*Psi=Psi}
\end{equation}
is factorized into the direct product of the projectors in each
space-time direction:
\begin{equation}
\ket{\Psi_C}=\ket{N}^0\otimes\cdots\otimes\ket{N}^{p}
\otimes\ket{D}^{p+1}\otimes\cdots\otimes\ket{D}^{25} ,
\end{equation}
where $\ket{N}^\mu$ and $\ket{D}^\mu$ are the translationally
invariant and the lump projectors for the direction $\mu$.
And the tachyon fluctuation mode around $\Psi_C$ with center-of-mass
momentum $k_\parallel$ in the directions $\mu=0,1,\cdots,p$
parallel to the brane is given by injecting $k_\parallel$ to the
midpoint of $\Psi_C$;
$\exp\left(ik_\parallel X^\parallel(\pi/2)\right)\ket{\Psi_C}$
\cite{fluctuations,note} (higher excitation modes in the longitudinal
directions are given by multiplying the tachyon mode by suitable
combinations of the creation operators $a_n^{\parallel\dagger}$
\cite{higher}).
Therefore, in order for the change of the boundary conditions in the
sense mentioned above is realized, it is at least necessary that the
string coordinate $X^\mu(\sigma)$ acting on the lump projector
$\ket{D}^\mu$ in the $\mu$-th direction vanishes at the endpoints (we
are assuming that the lump is located at $x^\mu=0$):
\begin{equation}
X^\mu(\sigma=0,\pi)\ket{D}^\mu =0 .
\label{eq:X(sigma=0,pi)ketD=0}
\end{equation}
Besides the string coordinate $X^\mu(\sigma)$, its conjugate momentum
$P_\mu(\sigma)$ should also satisfy the Dirichlet BC on
$\ket{D}^\mu$. However, we must be careful about the string parameter
$\sigma$.
It may happen that $P_\mu(\sigma)\ket{D}^\mu$ diverges at the
endpoints $\sigma=0$ and $\pi$ although the Dirichlet BC
\eqref{eq:X(sigma=0,pi)ketD=0} for $X^\mu(\sigma)$ is realized.
In this case we must suitably choose a new string parameter
$s=s(\sigma)$ in the neighborhood of each of the endpoints in such a
way that both $X^\mu(\sigma(s))\ket{D}^\mu$ and $P_\mu(s)\ket{D}^\mu$
have a regular series expansion in powers of $s$
(we assume that $s=0$ corresponds to the endpoint).
Here, $P_\mu(s)$ with an argument $s$ denotes the conjugate momentum
associated with the new string parameter;
$P_\mu(s)=(d\sigma/ds)P_\mu(\sigma(s))$.
We should examine whether the Dirichlet BC for the new conjugate
momentum
\begin{equation}
P_\mu(s=0)\ket{D}^\mu=0 ,
\label{eq:P(s=0)ketD=0}
\end{equation}
is satisfied or not.

We carry out the test of the Dirichlet BC
\eqref{eq:X(sigma=0,pi)ketD=0} and \eqref{eq:P(s=0)ketD=0} on the
one-parameter family of lump solutions given in \cite{BCFT,Muk}
using the boundary CFT technique.
We find that the condition \eqref{eq:X(sigma=0,pi)ketD=0} for
$X^\mu$ is satisfied only at a special value of the parameter.
The condition \eqref{eq:P(s=0)ketD=0} for the conjugate
$P_\mu$, however, cannot be satisfied even at that value of
the parameter.
The same test of the Dirichlet BC is done also for another
one-parameter family of lump projector obtained in \cite{solutions} by
the algebraic technique to lead to a negative result.
Therefore, our conclusion is that the Dirichlet BC cannot be
completely satisfied in the sense of \eqref{eq:X(sigma=0,pi)ketD=0}
and \eqref{eq:P(s=0)ketD=0} at lease for the known lump solutions in
VSFT.
As a byproduct of our analysis of the boundary conditions, we find
that, contrary to the previous expectation, the one-parameter family
of lump solutions given in \cite{BCFT,Muk} and that given in
\cite{solutions} are different ones.

On the other hand, it has been known that the matter projectors have
peculiar properties at the string midpoint $\sigma=\pi/2$: for
example, we have $X^\mu(\pi/2)\ket{D}^\mu=0$ for the lump projector
\cite{singular}.\footnote{
See also \cite{half,GTone,GTtwo,FuruOku} for issues concerning the
midpoint in VSFT.
}
This suggests that the midpoint could be interpreted
as a kind of string endpoint.
We examine the midpoint behavior of the string coordinate and its
conjugate acting on the projectors to find that the Dirichlet
(Neumann) condition in the above sense is indeed satisfied for the
lump (translationally invariant) projectors.
Physical interpretation of this result, however, is still unclear.

In the above analyses of the boundary and the midpoint behaviors,
we implicitly assume that the original oscillators $(a_n,a_n^\dagger)$
of \eqref{eq:X} and \eqref{eq:P} are the fundamental ones and study
the behavior of their coefficient functions.
However, we can define a new set of oscillators which is different
from $(a_n,a_n^\dagger)$ and in a sense more natural in discussing the
behaviors of the string coordinates and their conjugates acting on
projectors. The new annihilation operators are defined to annihilate a
projector instead of the Fock vacuum of the string modes.
We examine how the boundary and the midpoint behaviors are
changed if we adopt the new set of oscillators as the fundamental
ones.

The organization of the rest of this paper is as follows.
In sec.\ \ref{sec:Lump}, we carry out the analysis of the boundary and
the midpoint behaviors of the string coordinate and its conjugate
acting on a lump projector.
In sec.\ \ref{sec:NeumannSol}, we examine the midpoint behavior of the
translationally invariant projector.
In sec.\ \ref{sec:anotherlump}, we discuss the disagreement of the two
kinds of lump projectors by making use of the boundary conditions.
Then, in sec.\ \ref{sec:b}, we repeat the analysis of the midpoint
behavior for the translationally invariant projector by taking the new
oscillators.
Finally in sec.\ \ref{sec:final}, we summarize the paper and discuss
further problems. In the appendices, various technical details
used in the text are explained.

\section{Lump projector}
\label{sec:Lump}

As stated in sec.\ \ref{sec:Intro}, we would like to study the
boundary and the midpoint behaviors of the string coordinate
$X(\sigma)$ and its conjugate $P(\sigma)$ acting on the VSFT lump
projectors.
As a concrete example, let us first consider the lump solution
proposed in \cite{BCFT} using the boundary CFT technique. Its explicit
expression in the oscillator representation is given up to the overall
normalization by \cite{Muk}
\begin{equation}
\Muk=\int_{-\infty}^\infty\!dp\Mukp{p} ,
\label{eq:Muk}
\end{equation}
with
\begin{equation}
\Mukp{p}=\exp\!\left(
-\frac12\sum_{m,n=1}^\infty Q_{mn}a_m^\dagger a_n^\dagger
+p\sum_{n=1}^\infty \mL_n a_n^\dagger-\frac12 \mB\,p^2
\right)\ket{p} .
\label{eq:Mukp}
\end{equation}
Here and in the following, we omit the Lorentz index $\mu$ associated
with the projector and the oscillators in it.
In \eqref{eq:Mukp}, $\ket{p}$ is the eigenstate of the center-of-mass
momentum $\wh{p}$ satisfying
\begin{equation}
\wh{p}\ket{p}=p\ket{p},\quad
\wh{x}\ket{p}=-i\Drv{}{p}\ket{p} ,
\end{equation}
and $Q_{mn}$, $\mL_n$ and $\mB$ are given
as follows:
\begin{align}
Q_{mn}&=\frac{1}{\sqrt{mn}}\ooint{z}{z=0}\frac{h'(z)}{z^m}
\ooint{w}{w=0}\frac{h'(w)}{w^n}
\frac{1}{\left(h(z)-h(w)\right)^2}
\frac{h(z)h(w)-\tz^2}{\sqrt{\left(\tz^2-h(z)^2\right)
\left(\tz^2-h(w)^2\right)}} ,
\label{eq:Q}
\\
\mL_n &=\sqrt{\frac{2}{n}}\ooint{z}{z=0}
\frac{h'(z)}{z^n}\frac{\tz}{h(z)\sqrt{\tz^2-h(z)^2}} ,
\label{eq:mL}
\\
\mB &=2\,\ln\!\left(2 \tz\right) ,
\label{eq:mB}
\end{align}
where the function $h(z)$ and its derivative are
\begin{equation}
h(z)=\arctan z=\frac{1}{2i}\ln\frac{1+iz}{1-iz}, \quad
h'(z)=\frac{1}{1+z^2} .
\label{eq:h(z)}
\end{equation}
Note that the matrix $Q_{mn}$ and the vector $\mL_n$ are both
twist-even ones satisfying $CQC=Q$ and $C\bm{\mL}=\bm{\mL}$ with
$C_{mn}=(-1)^n\delta_{mn}$ being the twist matrix.
The present solution $\Muk$ carries a real parameter $\tz$ which
specifies the position of the twist operators in its BCFT
construction.\footnote{
Our parameter $t$ is equal to $t_0$ in \cite{Muk}.
}
The value of $\tz$ is restricted to $\tz\ge\pi/4$.

Now letting $X(\sigma)$ \eqref{eq:X} and $P(\sigma)$ \eqref{eq:P}
act on the solution $\Muk$ to express the result without using the
annihilation operators, we get
\begin{align}
X(\sigma)\Muk &=i\int\! dp\left(A(\sigma)\,p
- \sum_{n=1}^\infty J_n(\sigma)a_n^\dagger\right)\Mukp{p} ,
\label{eq:X(sigma)Muk}
\\
P(\sigma)\Muk &=\frac{1}{\pi}\int\! dp\left(B(\sigma)\,p
- \sum_{n=1}^\infty K_n(\sigma)a_n^\dagger\right)\Mukp{p} ,
\label{eq:P(sigma)Muk}
\end{align}
where the coefficient functions $A(\sigma)$, $J_n(\sigma)$,
$B(\sigma)$ and $K_n(\sigma)$ are
\begin{align}
A(\sigma)&=\bm{\mL}\cdot\bm{\z}(\sigma)-\mB ,
\label{Adef}\\
\bm{J}(\sigma)&=\left(Q+1\right)\bm{\z}(\sigma)-\bm{\mL} ,
\label{Jdef}\\
B(\sigma)&=\bm{\mL}\cdot\bm{\y}(\sigma)+1 ,
\label{Bdef}\\
\bm{K}(\sigma)&=\left(Q-1\right)\bm{\y}(\sigma) .
\label{Kdef}
\end{align}
Using the integral representations \eqref{eq:Q} and \eqref{eq:mL},
we can carry out the infinite summations (i.e., the multiplications
between matrices and vectors) in
\eqref{Adef}--\eqref{Kdef} to find that they are expressed in terms
of two kinds of functions, $C(\sigma)$ and $I_n(\sigma)$:
\begin{align}
C(\sigma)&=\frac12\ln\frac{
\tz -\sqrt{\tz^2-h(e^{i\sigma-\epsilon})^2}}
{\tz +\sqrt{\tz^2-h(e^{i\sigma-\epsilon})^2}} ,
\label{eq:Cdef}\\
I_n(\sigma)&=\frac{1}{\sqrt{2n}}\oint\limits_{z=0}\frac{dz}{2\pi i}
\frac{h'(z)}{z^n}
\frac{1}{h(e^{i\sigma-\epsilon})-h(z)}
\sqrt{\frac{\tz^2-h(e^{i\sigma-\epsilon})^2}{\tz^2-h(z)^2}} ,
\label{eq:Idef}
\end{align}
where $\epsilon$ is a positive infinitesimal and its origin is
the regularization factor $e^{-n\epsilon}$ multiplying $\z_n(\sigma)$
and $\y_n(\sigma)$.
In fact, the following relations hold (see appendix \ref{app:CI} for
their derivation):
\begin{align}
A(\sigma)&=C(\sigma)+C(-\sigma)=2 \Re C(\sigma),
\label{eq:AbyC}\\
J_n(\sigma)&=I_n(\sigma)+I_n(-\sigma)=2 \Re I_n(\sigma),
\label{eq:JbyI}\\
B(\sigma)&=\frac{1}{2i}\drv{}{\sigma}
\Bigl[C(\sigma)-C(-\sigma)\Bigr]=\drv{}{\sigma}\Im C(\sigma),
\label{eq:BbyC}\\
K_n(\sigma)&=\frac{1}{2i}\drv{}{\sigma}
\Bigl[I_n(\sigma)-I_n(-\sigma)\Bigr]
=\drv{}{\sigma}\Im I_n(\sigma).
\label{eq:KbyI}
\end{align}
Note that
\begin{equation}
C(\sigma)^*=C(-\sigma),\qquad
I_n(\sigma)^*=I_n(-\sigma) .
\end{equation}

Having finished the preparation, we shall proceed to the study of the
boundary and the midpoint behaviors of \eqref{eq:X(sigma)Muk} and
\eqref{eq:P(sigma)Muk}.
Concretely, we shall examine the behaviors of the functions
$A(\sigma)$, $J_n(\sigma)$, $B(\sigma)$ and $K_n(\sigma)$ which are
the coefficients of $p$ and $a_n^\dagger$.
For the sake of the easiness of explanation, we shall first look at
the midpoint behaviors.

\subsection{Midpoint behavior}
\label{subsec:Mukmid}

It has been known that the lump projectors $\ket{D}$ of VSFT share
the property $X(\pi/2)\ket{D}=0$ \cite{singular}.
Although the implication of this property is still controversial, it
suggests that the midpoint could be interpreted as the endpoint.
Here we shall study this property in more detail for the present
$\Muk$ \eqref{eq:Muk} for which the explicit calculation is possible.
In particular, we are interested in the behavior of both
$X(\sigma)\Muk$ and $P(\sigma)\Muk$ near the midpoint.

First, it is easily seen from $h(e^{\pm i\pi/2})=\pm i\infty$ that
$C(\pi/2)$ and $I_n(\pi/2)$ are both pure-imaginary and hence
we have $A(\pi/2)=J_n(\pi/2)=0$, implying that
\begin{equation}
X\!\left(\frac{\pi}{2}\right)\Muk=0 .
\label{eq:X(pi/2)Muk=0}
\end{equation}
For studying the behaviors near the midpoint, it is convenient to
express $h\!\left(e^{\pm i\sigma}\right)$ in terms of a new real
parameter $s\, (>0)$ as
\begin{align}
h\!\left(e^{\pm i\sigma}\right)&=
\begin{cases}
\ds \frac{\pi}{4}\pm\frac{i}{2s} ,
&\ds \left(0\le\sigma < \frac{\pi}{2}\right) ,
\\[5mm]
\ds -\frac{\pi}{4}\pm\frac{i}{2s} ,
&\ds \left(\frac{\pi}{2}<\sigma\le\pi\right) .
\end{cases}
\label{eq:s}
\end{align}
The explicit relation between $\sigma$ and $s$ is
\begin{equation}
\frac{1}{s}=\ln\frac{1+\sin\sigma}{\abs{\cos\sigma}}
\underset{\sigma\sim\frac{\pi}{2}}{\simeq}
\ln\frac{2}{\abs{(\pi/2)-\sigma}} ,
\label{eq:sbysigma}
\end{equation}
and the point $s=0$ correspond to the midpoint $\sigma=\pi/2$, while
$s=\infty$ to the endpoints $\sigma=0$ and $\pi$.
Then, it is easily seen that both $C(\sigma)$ \eqref{eq:Cdef} and
$I_n(\sigma)$ \eqref{eq:Idef} have regular series expansions in powers
of $s$ around $s=0$, and that their real (imaginary) parts are odd
(even) functions of $s$. For example, we have
\begin{equation}
C(\sigma(s))=\frac{i\pi}{2}-2\tz s-i\fugo\!\left(
\frac{\pi}{2}-\sigma\right)\pi\tz s^2
+\left(\frac{\pi^2}{2}+\frac43\tz^2\right)\tz s^3
+O\!\left(s^4\right) ,
\end{equation}
where $\fugo(x)=1$ ($-1$) when $x>0$ ($x<0$).
Therefore, $X(\sigma(s))\Muk$ has an expansion in odd powers of $s$,
which is consistent with the Dirichlet condition
\eqref{eq:X(pi/2)Muk=0}. Symbolically we have
\begin{equation}
X(\sigma(s))\Muk=s + s^3 + s^5 +\ldots .
\label{eq:X(s)Muksymbl}
\end{equation}
This implies that the natural string parameter near the midpoint is
not the original one $\sigma$ but rather is $s$.
Therefore, as the momentum variable conjugate to
$X(s)\equiv X(\sigma(s))$ we should take $P(s)$ which is related to the
original $P(\sigma)$ by\footnote{
We use the same symbol for the operators related by a conformal
transformation and distinguish them by their arguments.
}
\begin{equation}
P(s)=\drv{\sigma}{s}\,P(\sigma) ,
\label{eq:P(s)}
\end{equation}
in order to satisfy
\begin{equation}
\left[X(s),P(s')\right]=i\,\delta(s-s') .
\label{eq:[X(s),P(s')]=idelta}
\end{equation}
For this $P(s)$, $P(s)\Muk$ is given by \eqref{eq:P(sigma)Muk} with
$B(\sigma)$ and $K_n(\sigma)$ replaced with
\begin{equation}
B(s)=\drv{\sigma}{s}\,B(\sigma)=\drv{}{s}\Im C(\sigma),
\quad
K_n(s)=\drv{\sigma}{s}\,K_n(\sigma)=\drv{}{s}\Im I_n(\sigma) ,
\end{equation}
respectively. Recalling that $\Im C(\sigma)$ and $\Im I_n(\sigma)$
are even in $s$, we have symbolically
\begin{equation}
P(s)\Muk=s + s^3 + s^5 +\ldots .
\label{eq:P(s)Muksymbl}
\end{equation}
Namely, $P(s)\Muk$ also satisfies the Dirichlet condition at the
midpoint.
However, $P(\sigma)\Muk$ with the original string parameter $\sigma$
is {\em divergent} at the midpoint.
This is seen from
\begin{equation}
\drv{s}{\sigma}=-\frac{s^2}{\cos\sigma}
\underset{s\sim 0}{\simeq} -\frac12\,s^2 e^{1/s}\to\infty ,
\end{equation}
obtained from \eqref{eq:sbysigma}.
We shall discuss the implications of our result on the midpoint
behaviors in sec.\ \ref{sec:final}.

\subsection{Endpoint behavior}

Next let us study the behaviors of $X(\sigma)\Muk$ and
$P(\sigma)\Muk$ near the endpoints $\sigma=0, \pi$, which are directly
related to the change of the boundary conditions
(in the following we consider $\sigma=0$).
The endpoint behaviors of the basic functions $C(\sigma)$
\eqref{eq:Cdef} and $I_n(\sigma)$ \eqref{eq:Idef}
are completely different between the cases of $\tz>\pi/4$ and
$\tz=\pi/4$ since we have $h(1)=\pi/4$:
$C(\sigma)$ and $I_n(\sigma)$ have regular Taylor expansions in powers
of $\sigma$ for $\tz>\pi/4$, while their expansions are in odd powers
of $\sqrt{\sigma}$ (in particular, we have $C(0)=I_n(0)=0$) when
$\tz=\pi/4$. Explicitly, using the expansion
\begin{equation}
h\!\left(e^{i\sigma}\right)=\frac{\pi}{4}
+i\left(\frac{\sigma}{2}+\frac{\sigma^3}{12}
+O\left(\sigma^5\right)\right) ,
\end{equation}
we have
\begin{equation}
C(\sigma)=
\begin{cases}
\ds
\ln\frac{\pi/4}{\tz +\sqrt{\tz^2-(\pi/4)^2}}
+\frac{2i\tz}{\pi\sqrt{\tz^2-(\pi/4)^2}}\,\sigma
\\
\ds
\hspace{3.9cm}
+\frac{2\tz\left(\tz^2-(\pi^2/8)\right)}{
\pi^2\left(\tz^2-(\pi/4)^2\right)^{3/2}}\,\sigma^2
+O\!\left(\sigma^3\right) ,
&\ds \left(\tz>\frac{\pi}{4}\right) ,
\\[5mm]
\ds
-2\,e^{-i\pi/4}\sqrt{\frac{\sigma}{\pi}}
+\frac53\,e^{i\pi/4}\left(\frac{\sigma}{\pi}\right)^{3/2}
+O\!\left(\sigma^{5/2}\right) ,
&\ds \left(\tz=\frac{\pi}{4}\right) ,
\end{cases}
\label{eq:expandCend}
\end{equation}
and
\begin{align}
&I_n(\sigma)=\frac{1}{\sqrt{2n}}\oint\limits_{z=0}\frac{dz}{2\pi i}
\frac{h'(z)}{z^n}\frac{1}{(\pi/4)-h(z)}
\nn\\
&\times
\begin{cases}
\ds
\sqrt{\frac{\tz^2-(\pi/4)^2}{\tz^2-h(z)^2}}\left\{
1-\frac{i\left(\tz^2-(\pi/4)h(z)\right)}{
2\left(\tz^2-(\pi/4)^2\right)\left((\pi/4)-h(z)\right)}
\,\sigma +O\!\left(\sigma^2\right)
\right\} ,
&\ds \left(\tz>\frac{\pi}{4}\right) ,
\\[5mm]
\ds
\frac{1}{\sqrt{(\pi/4)^2-h(z)^2}}\biggl\{
\frac{e^{-i\pi/4}}{2}\sqrt{\pi\sigma}
-\frac{e^{i\pi/4}}{4\pi^2}\frac{(3\pi/4)+h(z)}{
(\pi/4)-h(z)}\left(\pi\sigma\right)^{3/2}
+O\!\left(\sigma^{5/2}\right)
\biggr\} ,
&\ds\left(\tz=\frac{\pi}{4}\right) .
\end{cases}
\label{eq:expandIend}
\end{align}
This result implies that, for $\tz>\pi/4$, the original Neumann
BC of $X(\sigma)$ and $P(\sigma)$ is kept unchanged even if they act
on $\Muk$.
However, the boundary condition is changed when these coordinates act
on $\Mukpof$ at the special value $\tz=\pi/4$ which corresponds to
putting the twist operators at ends of the local coordinate in the
BCFT construction of the solution \cite{BCFT,Muk}.
Let us concentrate on the latter case in the rest of this subsection.
First, $X(\sigma)\Mukpof$ satisfies the Dirichlet BC at the endpoints:
\begin{equation}
X(0)\Mukpof=0 .
\label{eq:X(0)Mukpof=0}
\end{equation}
Next, for considering the boundary condition of the conjugate momentum
$P$, we have to specify the natural string parameter near the
endpoints as we did for the midpoint in the previous subsection.
Since \eqref{eq:expandCend} and \eqref{eq:expandIend} for
$\tz=\pi/4$ are expanded in odd powers of $\sqrt{\sigma}$,
the natural string parameter near the endpoint $\sigma=0$ is
$u=\sqrt{\sigma}$ rather than the original $\sigma$.
For $X(u)\equiv X(\sigma)$ we have symbolically
\begin{equation}
X(u)\Mukpof=u + u^3 + u^5 + \ldots\, .
\label{eq:X(u)Mukpof}
\end{equation}
However, the expansion of $P(u)=(d\sigma/du)P(\sigma)$ acting on
$\Mukpof$ is in even powers of $u$:
\begin{equation}
P(u)\Mukpof=1 + u^2 + u^4 + \ldots\, .
\label{eq:P(u)Mukpof}
\end{equation}
Therefore, even at the special point $\tz=\pi/4$, the boundary
condition of $P(u)\Mukpof$ remains Neumann although that of
$X(u)\Mukpof$ is changed to Dirichlet.
The change of the boundary conditions in the background of the lump
solution \eqref{eq:Muk} cannot be completely realized in the expected
manner even in the case $\tz=\pi/4$.

\section{Midpoint behavior of the Neumann projector}
\label{sec:NeumannSol}

We have seen in sec.\ \ref{subsec:Mukmid} that the string coordinate
and its conjugate momentum acting on the lump projector \eqref{eq:Muk}
perfectly satisfy the Dirichlet condition at the midpoint.
In this section, we shall carry out the same kind of midpoint analysis
for the translationally
invariant Neumann projector $\ket{N}$ of VSFT \cite{solutions}, which
is given up to normalization by
\begin{equation}
\ket{N}=\exp\!\left(-\frac12\sum_{m,n=1}^\infty
S_{mn}a_m^\dagger a_n^\dagger\right)\ket{p=0} ,
\label{eq:ketN}
\end{equation}
where the matrix $S_{mn}$ has the following integral representation:
\begin{equation}
S_{mn}=\frac{1}{\sqrt{mn}}\ooint{z}{z=0}\frac{h'(z)}{z^m}
\ooint{w}{w=0}\frac{h'(w)}{w^n}
\frac{1}{\left(h(z)-h(w)\right)^2} .
\label{eq:S}
\end{equation}
Since $S_{mn}$ is equal to the $\tz\to\infty$ limit of $Q_{mn}$
\eqref{eq:Q} for the lump projector, i.e.,
$S_{mn}=\lim_{\tz\to\infty}Q_{mn}$,
most of the necessary formulas are obtained by simply taking the limit
in the corresponding equations in sec.\ \ref{sec:Lump} for the lump
projector.
We have
\begin{align}
X(\sigma)\ket{N}&=\left(\wh{x}-i\sum_{n=1}^\infty
J_n^N(\sigma)a_n^\dagger\right)\ket{N} ,
\label{eq:X(sigma)|N>}
\\
P(\sigma)\ket{N}&=-\frac{1}{\pi}\sum_{n=1}^\infty
K_n^N(\sigma)a_n^\dagger\ket{N} ,
\label{eq:P(sigma)|N>}
\end{align}
where $J_n^N(\sigma)$ and $K_n^N(\sigma)$ are given by
\begin{align}
\bm{J}^N(\sigma)&=\left(S+1\right)\bm{\z}(\sigma)
=\lim_{\tz\to\infty}\left(\bm{J}(\sigma)+\bm{\mL}\right) ,
\label{eq:bmJ^N}
\\
\bm{K}^N(\sigma)&=\left(S-1\right)\bm{\y}(\sigma)
=\lim_{\tz\to\infty}\bm{K}(\sigma) .
\label{eq:bmK^N}
\end{align}
Taking the $\tz\to\infty$ limit in \eqref{eq:JbyI} and
\eqref{eq:KbyI}, we obtain
\begin{align}
J_n^N(\sigma)&=2\Re I_n^N(\sigma)-\sqrt{2}\,t_n ,
\\
K_n^N(\sigma)&=\drv{}{\sigma}\Im I_n^N(\sigma) ,
\end{align}
where $I_n^N(\sigma)$ given by
\begin{equation}
I_n^N(\sigma)=\lim_{\tz\to\infty}I_n(\sigma)
=\frac{1}{\sqrt{2n}}\oint\limits_{z=0}\frac{dz}{2\pi i}
\frac{h'(z)}{z^n}\frac{1}{h(e^{i\sigma-\epsilon})-h(z)} ,
\label{eq:I^N}
\end{equation}
and $t_n$ is the vector appearing in the fluctuation modes
around the solution $\ket{N}$ \cite{fluctuations}:
\begin{equation}
\lim_{\tz\to\infty}\mL_n
=\sqrt{\frac{2}{n}}\ooint{z}{z=0}
\frac{h'(z)}{z^n h(z)}
=-\sqrt{2}\,t_n .
\label{eq:t_n}
\end{equation}
The natural string parameter in the present case is again $s$ of
\eqref{eq:s}. However, contrary to the case of $I_n(\sigma)$, the
present $I_n^N(\sigma)$ without
$\sqrt{\left(\tz^2-h(e^{i\sigma})^2\right)
/\left(\tz^2-h(z)^2\right)}$ has the property that
its real (imaginary) part is even (odd) in $s$.
Therefore, corresponding to \eqref{eq:X(s)Muksymbl} and
\eqref{eq:P(s)Muksymbl}, we have
\begin{align}
X(s)\ket{N}&=1 + s^2 + s^4 +\ldots ,
\\
P(s)\ket{N}&=1 + s^2 + s^4 +\ldots .
\end{align}
Namely, both $X(s)\ket{N}$ and $P(s)\ket{N}$ are subject to the
Neumann condition at the midpoint with respect to the new parameter
$s$:
\begin{equation}
\drv{}{s}X(s)\ket{N}\Bigr|_{s=0}
=\drv{}{s}P(s)\ket{N}\Bigr|_{s=0}=0 .
\label{eq:NB|N>atends}
\end{equation}
As for the boundary conditions of
$\left(X(\sigma),P(\sigma)\right)\ket{N}$ at the endpoints $\sigma=0$
and $\pi$, they remain Neumann since nothing singular happens at the
endpoints for the Neumann projector $\ket{N}$.

\section{Another lump projector}
\label{sec:anotherlump}

Besides the lump projector $\Muk$ which we discussed in sec.\
\ref{sec:Lump}, there is another kind of lump projector constructed in
\cite{solutions}. In this section we shall discuss the relationship
between the two lump projectors by making use of their boundary
behaviors.

First, let us summarize the lump solution of \cite{solutions}.
In its construction, they introduced the zero-mode oscillator
$(a_0,a_0^\dagger)$ which satisfies the usual commutation relation
$\bigl[a_0,a_0^\dagger\bigr]=1$:
\begin{equation}
a_0=\frac{\sqrt{b}}{2}\,\wh{p}-\frac{i}{\sqrt{b}}\,\wh{x} ,
\qquad
a^\dagger_0=\frac{\sqrt{b}}{2}\,\wh{p}+\frac{i}{\sqrt{b}}\,\wh{x} ,
\label{eq:a_0}
\end{equation}
where $b$ is an arbitrary positive parameter.
The lump projector $\SQb$ of \cite{solutions} is given up to
normalization by
\begin{equation}
\SQb=\exp\!\left(-\frac12\sum_{m,n=0}^\infty \Sb_{mn}
a_m^\dagger a_n^\dagger\right)\Omgb ,
\label{eq:SQb}
\end{equation}
where $\Omgb$ is the state annihilated by all $a_n$ including $a_0$:
\begin{equation}
a_n\Omgb=0 ,\quad (n\ge 0) .
\label{eq:Omgb}
\end{equation}
The coefficient matrix $\Sb$ in the exponent is given in quite a
similar manner to the algebraic construction of the translationally
invariant solution \eqref{eq:ketN} \cite{KosPot,solutions}.
We have to add a prime to all the matrices indicating that they are
extended ones including $n=0$.
First, $\Sb$ is related to another matrix $\Tb$ by
\begin{equation}
\Sb=\Cp\Tb ,
\label{eq:Sb=CpTb}
\end{equation}
with the (extended) twist-matrix $\Cp_{mn}=\delta_{mn}(-1)^n$
($m,n\ge 0$),
and $\Tb$ is given in terms of the primed Neumann matrix $\Mzb$ by
\begin{equation}
\Tb=\frac{1}{2\Mzb}{\left(1+\Mzb-\sqrt{(1-\Mzb)(1+3 \Mzb)}\right)} .
\label{eq:TbbyMzb}
\end{equation}
$\Mzb$ is expressed in terms of the unprimed Neumann matrix $\Mz$ and
the vectors $\bm{v}_0$ and $\bm{v}_1$ by\footnote{
Our convention of the Neumann matrices and vectors is given in
\cite{exact}.
}
\begin{equation}
\Mzb =\Pmatrix{(\Mzb)_{00} & (\Mzb)_{0n} \\ (\Mzb)_{m0} & (\Mzb)_{mn}}
=\Pmatrix{\ds
1-\frac23\frac{b}{\beta}
& \ds
\frac{\sqrt{2b}}{\beta}\bm{v}_0^\T
\\[10pt] \ds
\frac{\sqrt{2b}}{\beta}\bm{v}_0
& \ds
\Mz+\frac{1}{\beta}\left(
-3\bm{v}_0\bm{v}_0^\T + \bm{v}_1\bm{v}_1^\T\right)
} ,
\label{eq:Mzb}
\end{equation}
where $\beta$ is
\begin{equation}
\beta=2\Vzz +\frac{b}{2} ,
\label{eq:beta}
\end{equation}
with
\begin{equation}
\Vzz=\frac12\ln\!\left(\frac{3^3}{2^4}\right) .
\end{equation}
The solution \eqref{eq:SQb} has a free parameter $b$.
It was conjectured that $b$ is a kind of gauge
parameter \cite{solutions} although no proof has been given yet.

Using the relation
\begin{equation}
\ket{p}=\left(\frac{2\pi}{b}\right)^{-1/4}\exp\!\left(
-\frac{b}{4}\,p^2+\sqrt{b}\,p\,a_0^\dagger
-\frac12\bigl(a_0^\dagger\bigr)^2\right)
\ket{\Omega_b} ,
\end{equation}
the lump projector \eqref{eq:Muk} we considered in sec.\
\ref{sec:Lump} can also be put into the squeezed state form
\eqref{eq:SQb} using the zero-mode oscillators.
The corresponding $\Sb$ is
\begin{align}
&\Sb_{mn}=Q_{mn}-\frac{\mL_m \mL_n}{\mB+(b/2)} ,
\quad (m,n\ge 1) ,
\label{eq:SbmnbyQ}
\\
&\Sb_{n0}=\Sb_{0n}=-\frac{\sqrt{b}}{\mB+(b/2)}\,\mL_n ,
\quad (n\ge 1) ,
\label{eq:SbnzbyQ}
\\
&\Sb_{00}=\frac{\mB-(b/2)}{\mB+(b/2)} .
\label{eq:SbzzbyQ}
\end{align}

Now let us proceed to discussing the relation between the lump
solutions.
It was conjectured in \cite{Muk} that the two lump solutions,
\eqref{eq:Muk} with parameter $\tz$ and \eqref{eq:SQb} with $b$, are
the same thing; namely, they agree each other by giving $\tz$ as a
function of $b$ (or vice versa).
Numerical analysis for the relations
\eqref{eq:SbmnbyQ}--\eqref{eq:SbzzbyQ} was carried out
in \cite{Muk} to obtain results supporting the conjecture.
In this section, however, using the boundary conditions of $X(\sigma)$
acting on the two kinds of lump projectors, we present a negative
result for the agreement of the two.
More precisely, we shall show that the solution \eqref{eq:Muk}
at $\tz=\pi/4$ cannot coincide with the other solution \eqref{eq:SQb}
for any value of the parameter $b$.\footnote{
{}From the (numerical) agreement of the two $\Sb_{00}$, $\Sb_{00}$ for
\eqref{eq:SQb} given by \eqref{eq:TbbyMzb} and $\Sb_{00}$ of
\eqref{eq:SbzzbyQ}, it was claimed in \cite{Muk} that $\tz=\pi/4$
corresponds to $b\simeq 1.16$.
}

First recall \eqref{eq:X(0)Mukpof=0}, namely, that $X(0)$ acting on
the solution \eqref{eq:Muk} at $\tz=\pi/4$ vanishes.
On the other hand, for the solution \eqref{eq:SQb} we have
\begin{equation}
X(\sigma)\SQb=-i\left[
\bm{a}^{\prime\dagger}\cdot(\Sb +1)\bmzb(\sigma)\right]\SQb ,
\label{eq:X(sigma)SQb}
\end{equation}
where the primed vectors $\bmzb(\sigma)$ and $\bm{a}'$ are defined by
\begin{equation}
\bmzb(\sigma)=\Pmatrix{
\ds \frac{\sqrt{b}}{2}\\[12pt]
\ds \sqrt{\frac{2}{n}}\,\cos n\sigma}
=\Pmatrix{\ds \frac{\sqrt{b}}{2} \\[12pt]
\ds \bm{\z}(\sigma)} ,
\label{eq:bmzb}
\end{equation}
and $\bm{a}'=(a_0,a_1,a_2,\cdots)^\T$ including $a_0$, respectively.
Therefore, the condition for
\begin{equation}
X(0)\SQb=0 ,
\label{eq:X(0)SQb=0}
\end{equation}
is $(\Sb + 1)\bmzb(0)=0$,
which is equivalent to
\begin{equation}
\left(\Tb\pm 1\right)\frac{1\pm\Cp}{2}\bmzb(0)=0 ,
\label{eq:condbmzb}
\end{equation}
for the twist-even and odd parts of $\bmzb(0)$.
Since $\Tb$ is given by \eqref{eq:TbbyMzb} in terms of $\Mzb$, these
conditions are translated into the following two:
\begin{align}
\left(\Mzb+\frac13\right)\!\frac{1+\Cp}{2}\bmzb(0) &=0 ,
\label{eq:(Mzb+1/3)Zeven=0}
\\
(\Mzb-1)\frac{1-\Cp}{2}\bmzb(0) &=0 .
\label{eq:(Mzb-1)Zodd=0}
\end{align}
Let us consider the first condition \eqref{eq:(Mzb+1/3)Zeven=0}.
Its LHS is given, using \eqref{eq:Mzb} and that $\bm{v}_0$ is
twist-even, by
\begin{equation}
\left(\Mzb+\frac13\right)\frac{1+\Cp}{2}\bmzb(\sigma)
=\Pmatrix{
\ds
\frac{\sqrt{2b}}{\beta}\left(
\bm{v}_0\cdot\bm{\z}(\sigma)
+\frac{2\sqrt{2}}{3}\,\Vzz\right)
\\[7mm]
\ds
\frac{1}{\beta}\left(\frac{b}{\sqrt{2}}
-3\,\bm{v}_0\cdot\bm{\z}(\sigma)
\right)\bm{v}_0 +\left(\Mz+\frac13\right)
\frac{1+C}{2}\bm{\z}(\sigma)
} ,
\label{eq:(Mzb+1/3)Zeven}
\end{equation}
and hence \eqref{eq:(Mzb+1/3)Zeven=0} is reduced to the following
$b$-independent conditions for the unprimed quantities:
\begin{align}
\bm{v}_0\cdot\bm{\z}(\sigma)&=-\frac{2\sqrt{2}}{3}\,\Vzz ,
\label{eq:condzevenI}
\\
\left(\Mz+\frac13\right)\!\frac{1+C}{2}\bm{\z}(\sigma)
&= -\sqrt{2}\,\bm{v}_0 .
\label{eq:condzevenII}
\end{align}
However, the condition \eqref{eq:condzevenI} can be shown to be
invalid. In fact, the LHS of \eqref{eq:condzevenI} is calculated
using the integral representation of $\bm{v}_0$ to give
\begin{equation}
\bm{v}_0\cdot\bm{\z}(0)
= -\frac{2\sqrt{2}}{3}\Vzz +\frac{\sqrt{2}}{3}\ln 2 .
\label{eq:vzcdotz(0)}
\end{equation}
A derivation of \eqref{eq:vzcdotz(0)} is given in appendix
\ref{app:CI}.
Therefore, the boundary condition \eqref{eq:X(0)SQb=0} cannot hold for
any value of the parameter $b$, and hence $\Mukpof$ at $\tz=\pi/4$
cannot agree with $\SQb$ for any $b$.\footnote{
Eq.\ \eqref{eq:X(pi/2)Muk=0} at the midpoint holds also for $\SQb$
since we have $\left(\Sb+1\right)\bmzb(\pi/2)=0$ \cite{singular}.
}

\section{New oscillators}
\label{sec:b}

When we considered the boundary or midpoint behaviors of
$\left(X(\sigma),P(\sigma)\right)$ acting on the various projectors
$\ketproj$, we studied the coefficient functions of
$a_n^\dagger\ket{\mbox{Proj}}$.
However, there is a priori no reason why we have to look at the
coefficients of $a_n^\dagger\ketproj$ with the original
creation operator $a_n^\dagger$; in particular, $a_n^\dagger$ is not a
finite operator on the projectors in the sense that the norm of
$a_n^\dagger\ketproj$ is not finite even if $\ketproj$
is normalized.

In this subsection, we shall propose another way of considering the
boundary or midpoint behaviors.
This is to take, instead of the original $a_n$, the following
$b_n$:\footnote{The change of oscillators from $\bm{a'}$ to $\bm{b'}$
was considered previously by \cite{FKM} in the context of half-string
formalism \cite{half,GTone,GTtwo,FuruOku}.
We are grateful to the authors of \cite{FKM} for informing us of their
work.}
\begin{equation}
\bm{b}'=\frac{1}{\sqrt{1-\Sb^2}}
\left(\bm{a}'+\Sb\bm{a}'^\dagger\right) ,
\label{eq:b_n}
\end{equation}
with $\bm{b}'=\left(b_0,b_1,b_2,\cdots\right)^\T$ including $b_0$.
Here we are considering a lump projector ($\ketproj=\Muk$ or
$\SQb$) of the form \eqref{eq:SQb} in the $(a_0,a_0^\dagger)$
representation for the zero-mode.
The basic properties of $b_n$ are that it annihilates the solution,
\begin{equation}
b_n\ketproj=0 ,\quad
\left(n=0,1,2,\cdots\right) ,
\label{eq:b_nketSol=0}
\end{equation}
and that it is normalized:
\begin{equation}
\left[b_m,b_n^\dagger\right]=\delta_{mn} .
\label{eq:[b_m,b_n]=1}
\end{equation}
In terms of the new oscillator $(b_n,b_n^\dagger)$, the string
coordinate and its conjugate momentum are expanded as
\begin{align}
X(\sigma)&=i\left(\bm{b}'-\bm{b}'^\dagger\right)
\cdot\newbmz'(\sigma) ,
\\
P(\sigma)&=\frac{1}{\pi}\left(\bm{b}'+\bm{b}'^\dagger\right)
\cdot\newbmy'(\sigma) ,
\end{align}
where the function sets
$\newbmz'(\sigma)=\left(\newz'_n(\sigma)\right)$ and
$\newbmy'(\sigma)=\left(\newy'_n(\sigma)\right)$ are
defined by
\begin{align}
\newbmz'(\sigma)&=\left(\frac{1+\Sb}{1-\Sb}\right)^{1/2}
\bmzb(\sigma) ,
\label{eq:newbmz'}
\\
\newbmy'(\sigma)&=\left(\frac{1-\Sb}{1+\Sb}\right)^{1/2}
\bmyb(\sigma) ,
\label{eq:newbmy'}
\end{align}
with $\bmzb(\sigma)$ of \eqref{eq:bmzb} and
\begin{equation}
\bmyb(\sigma)=\Pmatrix{
\ds \frac{1}{\sqrt{b}}\\[12pt]
\ds \sqrt{\frac{n}{2}}\,\cos n\sigma}
=\Pmatrix{
\ds \frac{1}{\sqrt{b}}\\[12pt]
\ds \bm{\y}(\sigma)} .
\label{eq:bmyb}
\end{equation}
If we regard $(b_n,b_n^\dagger)$ as the basic oscillator,
it is the behavior of $\newbmz'(\sigma)$ and $\newbmy'(\sigma)$ at the
boundary or the midpoint that matters.
However, this analysis is not an easy task for the lump solutions
since there appear square roots of matrices in the definitions
\eqref{eq:newbmz'} and \eqref{eq:newbmy'}.

For the translationally invariant solution $\ket{N}$ \eqref{eq:ketN},
we can similarly define new oscillator $b_n$ annihilating $\ket{N}$.
It is given by the unprimed version of \eqref{eq:b_n}.
The corresponding mode functions are
\begin{align}
\newbmz(\sigma)&=\left(\frac{1+S}{1-S}\right)^{1/2}
\bm{\z}(\sigma) ,
\label{eq:newbmz}
\\
\newbmy(\sigma)&=\left(\frac{1-S}{1+S}\right)^{1/2}
\bm{\y}(\sigma) .
\label{eq:newbmy}
\end{align}
In this case of $\ket{N}$ we can carry out explicit calculations since
the eigenvalue problem of the Neumann matrices has been solved
completely \cite{spectroscopy}. We present the outline of the
calculation in appendix \ref{app:newbmzy}. In the neighborhood of the
midpoint $\sigma=\pi/2$, $\newz_n(\sigma)$ is expanded in odd powers
of $\sqrt{s}$ with $s$ defined by \eqref{eq:s}:
\begin{equation}
\newz_n(\sigma)=\mbox{const.}+\sqrt{\frac{2}{n}}
\oint\limits_{z=0}\frac{dz}{2\pi i}\frac{h'(z)}{z^{n}}
\Bigl[\sqrt{s}-h(z)\,s^{3/2}+O\!\left(s^{5/2}\right)\Bigr] .
\label{eq:expandnewz}
\end{equation}
On the other hand, $\newy_n(\sigma)$ is given as
\begin{equation}
\newy_n(\sigma)=\drv{}{\sigma}w_n(\sigma) ,
\label{eq:newybyw}
\end{equation}
with $w_n(\sigma)$ having also an expansion in odd powers of
$\sqrt{s}$:
\begin{equation}
w_n(\sigma)=\frac{1}{\sqrt{2n}}
\oint\limits_{z=0}\frac{dz}{2\pi i}\frac{h'(z)}{z^{n}}
\Bigl[\sqrt{s} +h(z)\,s^{3/2}+O\!\left(s^{5/2}\right)\Bigr] .
\label{eq:expandw}
\end{equation}
Recall that $I_n^N(\sigma)$ giving $\bm{J}^N(\sigma)$ \eqref{eq:bmJ^N}
and $\bm{K}^N(\sigma)$ \eqref{eq:bmK^N} has an expansion in integer
powers of $s$. In the present case the expansions have been changed to
those in terms of half an odd integer powers of $s$. This is due to
the square roots in \eqref{eq:newbmz} and \eqref{eq:newbmy} and that
the matrix $S$ has the eigenvalue $1$
(see appendix \ref{app:newbmzy}).

Eq.\ \eqref{eq:expandnewz} and \eqref{eq:expandw} imply that
the natural string parameter near the midpoint is $v=\sqrt{s}$
rather than $s$. The mode function $\newy_n(v)$ for the momentum
$P(v)$ corresponding to the string parameter $v$ is
\begin{equation}
\newy_n(v)=\drv{}{v}\Im w_n(\sigma)
=\frac{1}{\sqrt{2n}}
\oint\limits_{z=0}\frac{dz}{2\pi i}\frac{h'(z)}{z^{n}}
\Bigl[1 +3\,h(z) v^2+O\!\left(v^4\right)\Bigr] ,
\label{eq:expandnewy(v)}
\end{equation}
and it is subject to the Neumann condition at the midpoint (namely,
$(d/dv)\newy_n(v)=0$ at $v=0$).
However, $\newz_n(v)=\newz_n(\sigma)$ \eqref{eq:expandnewz} for
the coordinate $X(v)$ has an expansion in odd powers of $v$ and does
not satisfy the Neumann condition at the midpoint.
Namely, the midpoint behavior of the string coordinate acting on
$\ket{N}$ of \eqref{eq:ketN} depends on the choice of the basic
oscillators, $(a_n,a_n^\dagger)$ or $(b_n,b_n^\dagger)$.
It is our future problem to carry out similar analysis of the boundary
behavior for the lump projectors.

\section{Summary and discussions}
\label{sec:final}

In this paper we studied the endpoint and the midpoint behaviors of
the string coordinate and its conjugate momentum acting on the various
matter projectors in VSFT.
Our original expectation was that the Dirichlet BC is realized on the
lump projectors in spite of the fact that the string coordinates are
defined to obey the Neumann BC in VSFT.
Our findings are summarized as follows:
\begin{itemize}
\item The endpoint and the midpoint behaviors of the lump projector
$\Muk$ \eqref{eq:Muk} with a parameter $\tz$ and those of the
translationally invariant projector $\ket{N}$ \eqref{eq:ketN} are
summarized in table \ref{tbl:MukNXP}.
Here, we adopted the original oscillators $(a_n,a_n^\dagger)$ as the
basic ones; namely, we examined the behaviors of the coefficient
functions of $a_n^\dagger$ in $(X,P)\ketproj$.
\item The string coordinate acting on another lump projector $\SQb$
\eqref{eq:SQb} carrying a parameter $b$ does not satisfy the
Dirichlet BC at the endpoints for any $b$. This is a negative result
against the identification of the two lump projectors $\Muk$ and
$\SQb$ \cite{Muk}.
\item If, instead of the original oscillators $a_n$, we adopt as the
basic oscillators the new set $b_n$ \eqref{eq:b_n} which annihilates
the projector, the Neumann property of $X(\sigma)\ket{N}$ at the
midpoint no longer holds.
\end{itemize}
\begin{table}[htbp]
\begin{center}
\renewcommand{\arraystretch}{1.4}
\begin{tabular}{|c|c|c|c|c|}
\hline
&\multicolumn{2}{c|}{endpoint}&\multicolumn{2}{c|}{midpoint}\\
\cline{2-5}
& $X$ & $P$ & $X$ & $P$\\
\hline
$\ket{D_{\tz=\pi/4}}$ & D & N & D & D\\
\hline
$\ket{D_{\tz>\pi/4}}$ & N & N & D & D\\
\hline
$\ket{N}$ & N & N & N & N\\
\hline
\end{tabular}
\end{center}
\caption{The endpoint and the midpoint behaviors of the string
coordinate $X$ and its conjugate momentum $P$ acting on the
projectors $\Muk$ and $\ket{N}$ of VSFT. N and D denote the Neumann
and the Dirichlet conditions, respectively.
}
\label{tbl:MukNXP}
\end{table}

There remain many questions left unanswered.
First, as seen from table \ref{tbl:MukNXP}, our original expectation
on the dynamical change of the boundary conditions of $(X,P)$ from the
Neumann to the Dirichlet on the lump background is not completely
realized. Even in the case of $\Muk$ at $\tz=\pi/4$, the conjugate
momentum remains Neumann at the boundary although the string
coordinate obeys the Dirichlet BC.
This may imply that our expectation that change of the boundary
conditions occurs in $(X,P)\ketproj$ is too strong.
Another possibility would be that the change of the boundary conditions
is realized in the coefficient functions of other set of creation
operators in $(X,P)\ketproj$ than the original $a_n^\dagger$.
A candidate of such new set oscillators is $(b_n,b_n^\dagger)$ of
sec.\ \ref{sec:b}. In this paper, we saw only that the midpoint
behaviors of $(X,P)\ket{N}$ are different between the $a_n^\dagger$
and the $b_n^\dagger$ cases.
Carrying out a complete analysis of the boundary conditions using
$b_n^\dagger$ for the lump projectors $\Muk$ and $\SQb$ is one of our
future problems. Analysis of the eigenvalue problem of the primed
Neumann matrices in \cite{speczero} may be useful.
It is interesting if the new set of oscillators have special meaning
in the construction of the fluctuation modes around the solution in
VSFT \cite{fluctuations,higher}.

Our second problem is the interpretation of the midpoint behaviors of
$(X,P)\ketproj$. In this paper, we found that the midpoint obeys
perfectly the Neumann and the Dirichlet condition in the cases of the
translationally invariant and the lump projectors, respectively.
This may merely be a manifestation of the singular nature of VSFT with
purely ghost BRST operator \cite{singular,ghost}.
A more positive and radical interpretation of this phenomenon would be
that the midpoint is in fact the open string boundary. In particular
for $\ket{N}$, the property
$\int_0^{\pi/2}\!d\sigma P(\sigma)\ket{N}=0$
implying naively that we can split the left and the right halves of
the open string \cite{singular,projectors} seems to support our expectation.
Pursuing this possibility is also our interesting future problem.

\section*{Acknowledgments}
We would like to thank H.\ Ooguri and S.\ Teraguchi for valuable
discussions and comments.
S.\ M.\ is also grateful to theoretical particle physics group of
Department of Physics, Kyoto University for hospitality.
The work of H.\ H.\ was supported in part by the Grant-in-Aid for
Scientific Research (C) No.\ 15540268 from Japan Society for the
Promotion of Science (JSPS).
The work of S.\ M.\ was supported in part by the DOE grant
DE-FG03-92-ER40701.

\section*{Appendix}
\appendix

\section{Derivation of
(\ref{eq:AbyC})--(\ref{eq:KbyI}) and (\ref{eq:vzcdotz(0)})}
\label{app:CI}

In this appendix we briefly summarize the points in deriving
\eqref{eq:AbyC}--\eqref{eq:KbyI} and \eqref{eq:vzcdotz(0)}.
In carrying out the infinite summations in
\eqref{Adef}--\eqref{Kdef}, we multiply $\z_n(\sigma)$ and
$\y_n(\sigma)$ by the regularization factor $e^{-n\epsilon}$ and let
$\epsilon\to +0$ in the end. Let us consider, for example,
$J_n(\sigma)$ \eqref{Jdef}.
For this it is convenient to use another expression of $Q_{mn}$,
\begin{equation}
Q_{mn}=\sqrt{\frac{n}{m}}\ooint{z}{z=0}\frac{h'(z)}{z^{m}}
\ooint{w}{w=0}\frac{1}{w^{n+1}}\frac{1}{h(w)-h(z)}
\sqrt{\frac{\tz^2-h(w)^2}{\tz^2-h(z)^2}} ,
\end{equation}
which is obtained from \eqref{eq:Q} by integration by parts.
Then the first term on the RHS of \eqref{Jdef} with the
regularization is given by
\begin{align}
&\sum_{n=1}^\infty Q_{mn}\z_n(\sigma) e^{-n\epsilon}
\nn\\
&=\sqrt{\frac{2}{m}}\oint\limits_{z=0}\frac{dz}{2\pi i}
\frac{h'(z)}{z^m}\oint\limits_{C_w}\frac{dw}{2\pi i}
\frac{1}{h(w)-h(z)}\sqrt{\frac{\tz^2-h(w)^2}{\tz^2-h(z)^2}}
\frac{1}{2w}\sum_\pm\frac{1}{w e^{\epsilon\pm i\sigma}-1} .
\label{eq:calcJ}
\end{align}
The contour $C_w$ of the $w$-integration should be such
that encloses $w=e^{\pm i\sigma-\epsilon}$ but excludes the branch
points $w=\pm i$ of $h(w)$. The former requirement is due to the
convergence of the geometric series
$\sum_{n=1}^\infty \cos n\sigma e^{-n\epsilon}/w^{n+1}$.
Therefore, we should take residues at four points
$w=0$, $z$, and $e^{-\epsilon\pm i\sigma}$ in carry out the
$w$-integration (the contour of $z$-integration is a small one
enclosing $z=0$), which give $\mL_m$, $-\z_m(\sigma)$ and
$I_m(\pm\sigma)$, respectively, in the limit $\epsilon\to +0$.
This finishes the proof of \eqref{eq:JbyI}.
Derivations of \eqref{eq:BbyC} and \eqref{eq:KbyI} are quite similar.

Next, the derivation of \eqref{eq:AbyC} is a bit more involved.
The first term of $A(\sigma)$ \eqref{Adef} with the regularization
factor is written as
\begin{equation}
\sum_{n=1}^\infty\mL_n\z_n(\sigma)\,e^{-n\epsilon}
=\oint\limits_{C_z}\frac{dz}{2\pi i}
\left(\drv{}{z}\ln U(z)-\frac{1}{z}\right)
\sum_\pm\ln\!\left(\frac{z-e^{\pm i\sigma-\epsilon}}{z}\right) ,
\label{eq:calcA}
\end{equation}
where the contour $C_z$ is the same one as $C_w$ for \eqref{eq:calcJ},
and $U(z)$ is given by
\begin{equation}
U(z)=\frac{z}{h(z)}\left(\tz +\sqrt{\tz^2-h(z)^2}\right) .
\label{eq:U(z)}
\end{equation}
Note that $U(z)$ is regular inside $C_z$, in particular, at $z=0$.
By taking as $C_z$ the ones running just above and below the
logarithmic cut connecting $z=0$ and $e^{\pm i\sigma-\epsilon}$ for
each of the two terms in the summation $\sum_\pm$ in \eqref{eq:calcA},
we see that the contribution to \eqref{eq:calcA} of the $(d/dz)U(z)$
term is
$\sum_\pm\left(\ln U(0)-\ln U(e^{\pm i\sigma})\right)
=\mB+C(\sigma)+C(-\sigma)$,
while that of the $-1/z$ term vanishes.
Hence we get \eqref{eq:AbyC}.

Finally, we present the derivation of \eqref{eq:vzcdotz(0)}, which is
quite similar to that of \eqref{eq:AbyC} explained above. For this we
need the integration formula for the Neumann vector $\bm{v}_0$:
\begin{equation}
(\bm{v}_0)_n
=-\frac{1}{3\sqrt{n}}\oint\limits_{z=0}\!\frac{dz}{2\pi i}
\frac{f'(z)}{z^n}
\left[
\frac{2}{f(z)-1} -\frac{1}{f(z)-\omega}-\frac{1}{f(z)-\omega^*}
\right] ,
\label{eq:v0int}
\end{equation}
with
\begin{equation}
f(z)=\left(\frac{1+ iz}{1-iz}\right)^{2/3},
\quad
f'(z)=\frac{4 i}{3}\frac{f(z)}{1+z^2},
\quad
\omega=e^{2\pi i/3} .
\end{equation}
Then, $\bm{v}_0\cdot\bm{\z}(\sigma)$ with the regularization is
calculated as follows:
\begin{align}
\sum_{n=1}^\infty(\bm{v}_0)_n\z_n(\sigma) e^{-n\epsilon}
&=\frac{1}{3\sqrt{2}}\oint\limits_{C_z}\frac{dz}{2\pi i}
\left(\drv{}{z}\ln W(z) + \frac{2}{z}\right)
\sum_{\pm}\ln\left(\frac{z-e^{\pm i\sigma-\eps}}{z}\right)
\nn\\
&=\frac{1}{3\sqrt{2}}\sum_{\pm}\ln\frac{W(0)}{W(e^{\pm i\sigma})}
=-\frac{2\sqrt{2}}{3}\,\Vzz
-\frac{1}{3\sqrt{2}}\sum_{\pm}\ln W(e^{\pm i\sigma}) ,
\label{eq:calcvzcdotz(sigma)}
\end{align}
where $W(z)$ is defined by
\begin{equation}
W(z)=\frac{(f(z)-1)^2}{(-z^2)(f(z)-\omega)(f(z)-\omega^*)} ,
\label{eq:W(z)}
\end{equation}
and we have used that $W(0)=2^4/3^3$.
Putting $\sigma=0$ in \eqref{eq:calcvzcdotz(sigma)} and using that
$W(1)=1/2$, we obtain \eqref{eq:vzcdotz(0)}.

\section{ $\newbmz(\sigma)$ and $\newbmy(\sigma)$}
\label{app:newbmzy}

In this appendix we derive the series expansions \eqref{eq:expandnewz}
and \eqref{eq:expandw} of the functions
$\newbmz(\sigma)$ \eqref{eq:newbmz} and $\newbmy(\sigma)$
\eqref{eq:newbmy} near the midpoint.
We apply the technique developed in \cite{spectroscopy} and
\cite{ratio,kinetic} for the eigenvalue problem of the Neumann
coefficients of the translationally invariant solution.
Here we use bras and kets for the vectors by following
\cite{ratio,kinetic}.
Let $\ket{\kappa}$ be the eigenvector of the the infinite dimensional
symmetric matrix $K_1$ with components
$\left(K_1\right)_{mn}=-\sqrt{(m-1)m}\,\delta_{m-1,n}
-\sqrt{m(m+1)}\,\delta_{m+1,n}$
corresponding to the eigenvalue $\kappa$:
\begin{equation}
K_1\ket{\kappa}=\kappa\ket{\kappa} ,
\quad
\left(-\infty<\kappa<\infty\right) .
\end{equation}
The Neumann matrix $\Mz$ is expressed in terms of $K_1$ as
$\Mz=-\left(1+2\,\cosh\left(\pi K_1/2\right)\right)^{-1}$, and hence
$\ket{\kappa}$ is also the eigenvector of $T=SC$ given by
\eqref{eq:TbbyMzb} without primes:
\begin{equation}
T\ket{\kappa}=-e^{-(\pi/2)\abs{\kappa}}\ket{\kappa} .
\end{equation}
Other basic formulas concerning $\ket{\kappa}$ are
\begin{align}
\braket{n}{\kappa}&=
\frac{\sqrt{n}}{\kappa}\oint\limits_{z=0}
\frac{dz}{2\pi i}\frac{1}{z^{n+1}}\left(1- e^{-\kappa h(z)}\right) ,
\label{eq:<n|kappa>}
\\
\int_{-\infty}^\infty\frac{d\kappa}{\calN(\kappa)}
\ket{\kappa}\!\bra{\kappa}&=\bm{1} ,
\quad
\left(\calN(\kappa)=\frac{2}{\kappa}\sinh\frac{\pi\kappa}{2}
\right) ,
\\
C\ket{\kappa}&=-\ket{-\kappa} ,
\end{align}
where $\ket{n}$ ($n=1,2,3,\cdots$) is the vector with its $n$-th
component equal to one and all other components equal to zero.

The inner-products of $\ket{\kappa}$ with
the vectors $\ket{\z(\sigma)}$ \eqref{eq:z_n} and $\ket{\y(\sigma)}$
\eqref{eq:y_n} are calculated using \eqref{eq:<n|kappa>} to give
\begin{align}
\braket{\kappa}{\z(\sigma)}
&=\frac{\sqrt{2}}{\kappa}\Re\left(1- e^{-\kappa h(e^{i\sigma})}\right) ,
\\
\braket{\kappa}{y(\sigma)}
&=\frac{1}{\sqrt{2}\kappa}\drv{}{\sigma}
\Im\left(1- e^{-\kappa h(e^{i\sigma})}\right) .
\end{align}
Then, using the above formulas and
\begin{equation}
\left(\frac{1\pm S}{1\mp S}\right)^{1/2}\ket{\kappa}
=\frac{1\pm TC}{\sqrt{1-T^2}}\ket{\kappa}
=\frac{1}{\sqrt{1- e^{-\pi\abs{\kappa}}}}\Bigl(
\ket{\kappa} \pm e^{-(\pi/2)\abs{\kappa}}\ket{-\kappa}\Bigr) ,
\end{equation}
we obtain
\begin{align}
\newz_n(\sigma)&=
\int_{-\infty}^\infty\frac{d\kappa}{\calN(\kappa)}
\bra{n}\left(\frac{1+S}{1-S}\right)^{1/2}
\ket{\kappa}\!\braket{\kappa}{\z(\sigma)}
\nn\\
&=\sqrt{\frac{2}{n}}\Re\oint\limits_{z=0}
\!\frac{dz}{2\pi i}\frac{h'(z)}{z^{n}}
\Bigl\{L_1(h(z))-L_2(h(z))
-L_1\!\left(h(e^{i\sigma})+h(z)\right)
-L_2\!\left(h(e^{i\sigma})-h(z)\right)
\Bigr\},
\end{align}
where $L_1(h)$ and $L_2(h)$ are defined by
\begin{align}
L_1(h)&=\int_{0}^\infty\frac{d\kappa}{\kappa\,\calN(\kappa)}
\frac{1}{\sqrt{1-e^{-\pi\kappa}}}
\left(e^{-\kappa h}-e^{\kappa h}\right) ,
\label{eq:L1(h)}
\\
L_2(h)&=\int_{0}^\infty\frac{d\kappa}{\kappa\,\calN(\kappa)}
\frac{e^{-(\pi/2)\kappa}}{\sqrt{1-e^{-\pi\kappa}}}
\left(e^{-\kappa h}-e^{\kappa h}\right) .
\label{eq:L2(h)}
\end{align}
Similarly, $w_n(\sigma)$ of \eqref{eq:newybyw} is given by
\begin{equation}
w_n(\sigma)
=\frac{1}{\sqrt{2n}}\Im\oint\limits_{z=0}
\frac{dz}{2\pi i}\frac{h'(z)}{z^{n}}
\Bigl\{L_1(h(z))+L_2(h(z))
-L_1\!\left(h(e^{i\sigma})+h(z)\right)
+L_2\!\left(h(e^{i\sigma})-h(z)\right)
\Bigr\} .
\label{eq:w_n}
\end{equation}
The functions $L_1(h)$ and $L_2(h)$ are expressed in terms of the
gamma functions as
\begin{align}
L_1(h)&=-\frac{2\,\sin h}{\pi^{3/2}}\left\{
\Gamma\!\left(1+\frac{h}{\pi}\right)
\Gamma\!\left(\frac12-\frac{h}{\pi}\right)
+\Gamma\!\left(1-\frac{h}{\pi}\right)
\Gamma\!\left(\frac12+\frac{h}{\pi}\right)\right\} ,
\label{eq:L1byGamma}
\end{align}
and
\begin{align}
L_2(h)&=-\frac{2\cos h}{\pi^{3/2}}
\left\{
\Gamma\!\left(1+\frac{h}{\pi}\right)
\Gamma\!\left(\frac12-\frac{h}{\pi}\right)
-\Gamma\!\left(1-\frac{h}{\pi}\right)
\Gamma\!\left(\frac12+\frac{h}{\pi}\right)\right\} .
\label{eq:L2byGamma}
\end{align}
The expansions \eqref{eq:expandnewz} and \eqref{eq:expandw} around the
midpoint are obtained by using the definition \eqref{eq:s} of the
parameter $s$ and the Stirling formula for the gamma function.


\begin{thebibliography}{99}

\bibitem{descent}
A.~Sen,
``Descent relations among bosonic D-branes,''
Int.\ J.\ Mod.\ Phys.\ A {\bf 14}, 4061 (1999)
[arXiv:hep-th/9902105].

\bibitem{tachyon}
A.~Sen and B.~Zwiebach,
``Tachyon condensation in string field theory,''
JHEP {\bf 0003}, 002 (2000)
[arXiv:hep-th/9912249].

\bibitem{CSFT}
E.~Witten,
``Noncommutative Geometry And String Field Theory,''
Nucl.\ Phys.\ B {\bf 268}, 253 (1986).

\bibitem{HarKra}
J.~A.~Harvey and P.~Kraus,
``D-branes as unstable lumps in bosonic open string field theory,''
JHEP {\bf 0004}, 012 (2000)
[arXiv:hep-th/0002117].

\bibitem{MJMT}
R.~de Mello Koch, A.~Jevicki, M.~Mihailescu and R.~Tatar,
``Lumps and p-branes in open string field theory,''
Phys.\ Lett.\ B {\bf 482}, 249 (2000)
[arXiv:hep-th/0003031].

\bibitem{MSZ}
N.~Moeller, A.~Sen and B.~Zwiebach,
``D-branes as tachyon lumps in string field theory,''
JHEP {\bf 0008}, 039 (2000)
[arXiv:hep-th/0005036].

\bibitem{VSFT}
L.~Rastelli, A.~Sen and B.~Zwiebach,
``String field theory around the tachyon vacuum,''
Adv.\ Theor.\ Math.\ Phys.\  {\bf 5}, 353 (2002)
[arXiv:hep-th/0012251].

\bibitem{solutions}
L.~Rastelli, A.~Sen and B.~Zwiebach,
``Classical solutions in string field theory around the tachyon vacuum,''
Adv.\ Theor.\ Math.\ Phys.\  {\bf 5}, 393 (2002)
[arXiv:hep-th/0102112].

\bibitem{BCFT}
L.~Rastelli, A.~Sen and B.~Zwiebach,
``Boundary CFT construction of D-branes in vacuum string field
theory,''
JHEP {\bf 0111}, 045 (2001)
[arXiv:hep-th/0105168].

\bibitem{Muk}
P.~Mukhopadhyay,
``Oscillator representation of the BCFT construction of D-branes in
vacuum string field theory,''
JHEP {\bf 0112}, 025 (2001)
[arXiv:hep-th/0110136].

\bibitem{singular}
G.~W.~Moore and W.~Taylor,
``The singular geometry of the sliver,''
JHEP {\bf 0201}, 004 (2002)
[arXiv:hep-th/0111069].

\bibitem{anomaly}
H.~Hata and S.~Moriyama,
``Observables as twist anomaly in vacuum string field theory,''
JHEP {\bf 0201}, 042 (2002)
[arXiv:hep-th/0111034].

\bibitem{exact}
H.~Hata, S.~Moriyama and S.~Teraguchi,
``Exact results on twist anomaly,''
JHEP {\bf 0202}, 036 (2002)
[arXiv:hep-th/0201177].

\bibitem{fluctuations}
H.~Hata and T.~Kawano,
``Open string states around a classical solution in vacuum string
field theory,''
JHEP {\bf 0111}, 038 (2001)
[arXiv:hep-th/0108150].

\bibitem{note}
L.~Rastelli, A.~Sen and B.~Zwiebach,
``A note on a proposal for the tachyon state in vacuum string field
theory,''
JHEP {\bf 0202}, 034 (2002)
[arXiv:hep-th/0111153].

\bibitem{higher}
H.~Hata and H.~Kogetsu,
``Higher level open string states from vacuum string field theory,''
JHEP {\bf 0209}, 027 (2002)
[arXiv:hep-th/0208067].

\bibitem{half}
L.~Rastelli, A.~Sen and B.~Zwiebach,
``Half strings, projectors, and multiple D-branes in vacuum string
field theory,''
JHEP {\bf 0111}, 035 (2001)
[arXiv:hep-th/0105058].

\bibitem{GTone}
D.~J.~Gross and W.~Taylor,
``Split string field theory. I,''
JHEP {\bf 0108}, 009 (2001)
[arXiv:hep-th/0105059].

\bibitem{GTtwo}
D.~J.~Gross and W.~Taylor,
``Split string field theory. II,''
JHEP {\bf 0108}, 010 (2001)
[arXiv:hep-th/0106036].

\bibitem{FuruOku}
K.~Furuuchi and K.~Okuyama,
``Comma vertex and string field algebra,''
JHEP {\bf 0109}, 035 (2001)
[arXiv:hep-th/0107101].

\bibitem{KosPot}
V.~A.~Kostelecky and R.~Potting,
``Analytical construction of a nonperturbative vacuum for the open  bosonic
string,''
Phys.\ Rev.\ D {\bf 63}, 046007 (2001)
[arXiv:hep-th/0008252].

\bibitem{FKM}
E.~Fuchs, M.~Kroyter and A.~Marcus,
``Continuous half-string representation of string field theory,''
JHEP {\bf 0311}, 039 (2003)
[arXiv:hep-th/0307148].

\bibitem{spectroscopy}
L.~Rastelli, A.~Sen and B.~Zwiebach,
``Star algebra spectroscopy,''
JHEP {\bf 0203}, 029 (2002)
[arXiv:hep-th/0111281].

\bibitem{speczero}
B.~Feng, Y.~H.~He and N.~Moeller,
``The spectrum of the Neumann matrix with zero modes,''
JHEP {\bf 0204}, 038 (2002)
[arXiv:hep-th/0202176].

\bibitem{ghost}
D.~Gaiotto, L.~Rastelli, A.~Sen and B.~Zwiebach,
``Ghost structure and closed strings in vacuum string field theory,''
Adv.\ Theor.\ Math.\ Phys.\  {\bf 6}, 403 (2003)
[arXiv:hep-th/0111129].

\bibitem{projectors}
D.~Gaiotto, L.~Rastelli, A.~Sen and B.~Zwiebach,
``Star algebra projectors,''
JHEP {\bf 0204}, 060 (2002)
[arXiv:hep-th/0202151].

\bibitem{ratio}
K.~Okuyama,
``Ratio of tensions from vacuum string field theory,''
JHEP {\bf 0203}, 050 (2002)
[arXiv:hep-th/0201136].

\bibitem{kinetic}
K.~Okuyama,
``Ghost kinetic operator of vacuum string field theory,''
JHEP {\bf 0201}, 027 (2002)
[arXiv:hep-th/0201015].

\end{thebibliography}
\end{document}